# Room temperature single-crystal diffuse scattering and ab initio lattice dynamics in CaTiSiO$_5$

[1]M. J. Gutmann, [2]K. Refson, [3]M. v. Zimmermann, [4]I. P. Swainson, [5]A. Dabkowski, and [5]H. Dabkowska

[1]Rutherford Appleton Laboratory, ISIS Facility, Chilton Didcot, Oxfordshire OX11 0QX, United Kingdom

[2]Rutherford Appleton Laboratory, CSE Department, Chilton Didcot, Oxfordshire OX11 0QX, United Kingdom

[3]Hamburger Synchrotronstrahlungslabor HASYLAB at Deutsches Elektronen-Synchrotron DESY, D-22603 Hamburg, Germany

[4]Canadian Neutron Beam Centre, National Research Council, Chalk River Laboratories, Chalk River, Ontario, Canada KOJ 1JO.

[5]Brockhouse Institute of Materials Research and Department of Physics and Astronomy, McMaster University, Hamilton, Canada

**Abstract**

Single-crystal diffuse scattering data have been collected at room temperature on synthetic titanite using both neutrons and high-energy X-rays. A simple ball-and-springs model reproduces well the observed diffuse scattering, confirming its origin to be primarily due to thermal motion of the atoms. Ab initio phonons are calculated using density-functional perturbation theory and are shown to reproduce the experimental diffuse scattering. The observed diffuse X-ray and neutron scattering patterns are consistent with a summation of mode frequencies and displacement eigenvectors associated with the entire phonon spectrum, rather than with a simple, short-range static displacement. A band gap is observed between 600 and 700 cm$^{-1}$ with only two modes crossing this region, both associated with antiferroelectric Ti-O motion along *a*. One of these modes (of $B_u$ symmetry), displays a large LO-TO mode splitting (562 cm$^{-1}$ to 701.4 cm$^{-1}$) and has a dominant component coming from Ti-O bond-stretching and, thus, the mode-splitting is related to the polarisability of the Ti-O bonds along the chain direction. Similar mode-splitting is observed in piezo- and ferroelectric materials. The calculated phonon dispersion model may be of use to others in future to understand the phase transition at higher temperatures, as well as in the interpretation of measured phonon dispersion curves.

**I. Introduction**

Titanite, CaTiSiO$_5$, is a naturally occurring mineral. It is used as a source of TiO$_2$ for pigments and as an inexpensive gemstone for jewellery [1]. This mineral is also known as sphene, which is related to greek word for 'wedge', an often encountered shape for the naturally occurring form.

At room temperature, titanite crystallises in a monoclinic structure, space group P 1 $2_1$/a 1 with lattice parameters $a$ = 7.0619(13)Å, $b$ = 8.7085(16)Å, $c$ = 6.5562(11)Å, β=113.850(11)°. The structure is made of a highly interconnected polyhedral network. $TiO_6$ octahedra form linear chains along the *a*-direction, and display alternating tilts within and along the chains. These chains are linked by $SiO_4$ tetrahedra, with Ca occupying the remainder of the polyhedral network in irregular seven-fold coordinated oxygen cavities. At room temperature, the Ti are slightly displaced off-center along *a*, in opposite sense in neighbouring chains. This leads to an antiferroelectric distortion pattern with no net external ferroelectric moment. However, no effect was observed when applying an electric field along the *a* direction, leading to the conclusion that switching of Ti displacements was not possible in strong electric fields along *a* between room-temperature and 500 K for fields at least up to 35 kV/cm [2]. Details of the crystal structure have been extensively discussed in [1]. The main interest of previous studies on both the crystal structure and phonons has been a structural phase transition near 500 K to another monoclinic structure, spacegroup A2/a and a second anomaly above 850 K [2, 3, 4]. In this high-temperature phase, the Ti ions nominally lie in the center of the octahedra, albeit with a large displacement parameter. This led to the idea that the high-temperature, paraelectric phase is due to a superposition of short-range ordered domains of the room-temperature phase and this was shown to reproduce temperature-dependent streaks around the (4 0 2) reflection in a single crystal diffuse X-ray scattering experiment [5]. The second phase appearing above 850 K was shown to correspond to the appearance of the pure A2/a phase [2].

Phonon measurements to date have focused on the phase transition using Raman and infrared spectroscopy [3, 6]. Both techniques probe the phonons at the center of the Brillouin zone. Phonon dispersion curves have not been determined yet and no theoretical curves published in the literature. The former would be useful to further establish the nature of the phase transition being of an order-disorder type as suggested by diffraction studies and/or driven by phonon-softening. Compared to synthetic crystals, natural mineral specimens often contain impurities such as Fe and Al partially substituting for Ti, or actinides, such as U and Th, partially replacing Ca. For its ability to accommodate actinides, titanite was suggested as a possible mineral for storing radioactive waste [7]. The presence of decaying actinides induces self-radiation damage [4, 7]. Impurities modify the phonon spectra and were found to have a pronounced effect on the detectability of Ti-O bond stretching modes [7].

In this paper, we carry out a detailed study of the room temperature phase using a large scale reciprocal space survey of the diffuse scattering using both neutrons and synchrotron X-rays. The interpretation of the diffuse scattering is aided by Monte-Carlo simulations and detailed ab initio lattice dynamics calculations using density-functional perturbation theory (DFPT). This allows a detailed study of the phonon dynamics paving the way for future inelastic neutron or X-ray scattering studies of the dispersion curves.

## II. Experimental

A large single crystal of titanite was grown in air by the floating zone method using halogen lamps as a heating source. CaTiSiO$_5$ melts congruently in air at 1382 °C. Stoichiometric ceramic rods were pre-annealed at 1260 °C for 12 hours and the crystal growth was performed with the speed 1 mm/h and rotation of both rods at 25 rpm. Neutron single crystal data were collected at room temperature using the time-of-flight Laue diffractometer SXD at the ISIS pulsed neutron source. The large detector array together with the ability to use a wide band of incident wavelengths allows a large volume of reciprocal space to be mapped [8]. The bottom 17 mm of the crystal was illuminated, and both Bragg and diffuse scattering were simultaneously collected using eight orientations and an exposure time of the order of 20 hours per orientation. A second small crystallite was found present, however its diffuse scattering could not be discerned. Data were treated using the locally available SXD2001 software [9]. Synchrotron X-ray data were recorded at the high-energy X-ray diffraction beamline BW5 at DESY, HASYLAB, employing a Perkin-Elmer XRD1621 CCD detector. The detector was controlled using the program QXRD [10]. A crystal of 1 mm$^2$ size was mounted on a loop on a spindle and rotated through a range of 180° with an oscillation range of 0.25°. Four 5-second exposures were summed per frame during the 0.25° oscillation and a dark frame automatically subtracted. The same scan was repeated three times. Data were indexed using XDS [11] and select reciprocal space planes were extracted using our own in-house code. Structure refinements were carried out using the JANA2006 program [12].

## III. Results and initial diffuse modelling

The average structure was refined against the neutron and X-ray data simultaneously and agrees with previous studies [1], confirming the antiferroelectric distortion pattern of off-center displacements of Ti along the TiO$_6$ polyhedral chains (Figure 1). Details are reported in Tables 1 and 2. Eight sections in reciprocal space were extracted from the neutron and X-ray data (Figures 2 and 3). The diffuse scattering appears rather weak and becomes more pronounced towards increasing Q. Features broadly take the shape of concentric trapezoids propagating along $b^*$, hence forming planar features along that direction. Several layers also display short streaks, reminiscent of what has been described in [5].

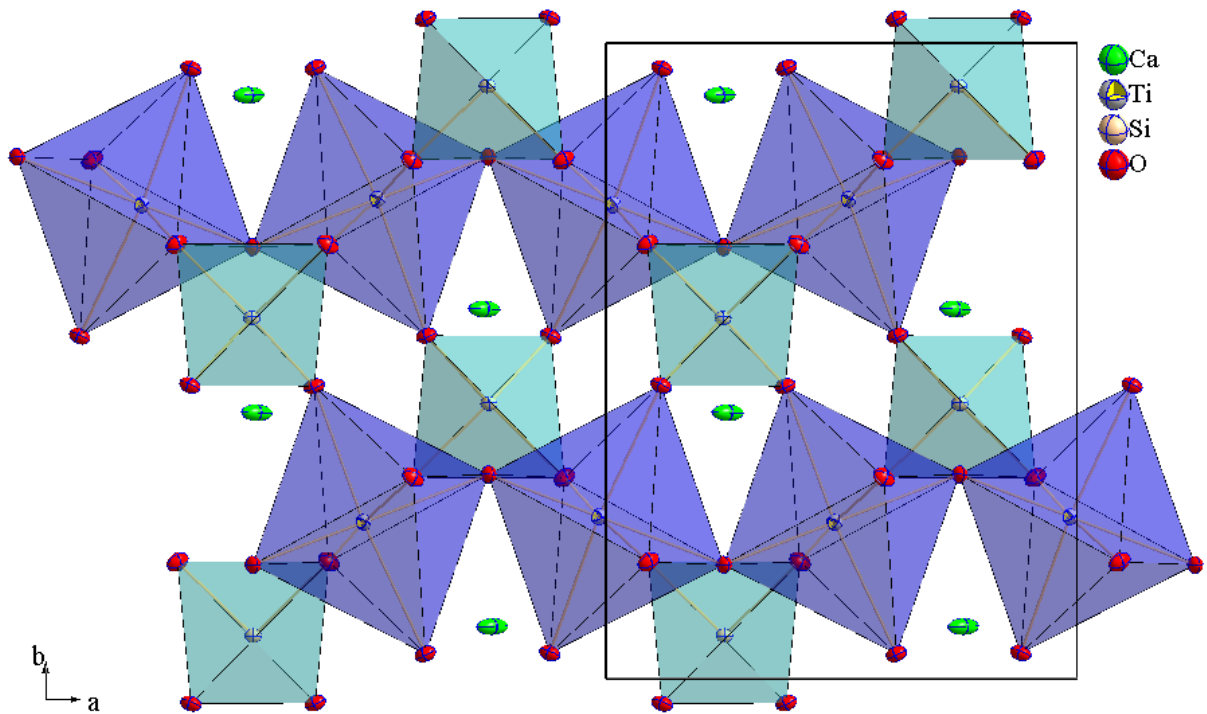

Figure 1 (Color online): Crystal structure of $CaTiSiO_5$ from the joint X-ray and neutron refinement. The unit cell content is expanded according to $2a \times b \times c$. Thermal ellipsoid are shown corresponding to 50% probability level. Chains of $TiO_6$ octahedra (blue) and linking $SiO_4$ tetrahedra (green) are highlighted, and the unit cell is indicated.

Table 1: Neutron and X-ray data collection data collection and refinement

| Formula | $CaTiSiO_5$ |
|---|---|
| | Neutron / X-ray |
| Formula weight | 196.03 |
| Temperature, K | 300(2) / RT |
| Wavelength range, Å | 0.35 - 10.0 / 0.12398 |
| Crystal system | Monoclinic |
| Space group | $P\ 1\ 2_1/a\ 1$ |
| Unit cell parameters, Å | $a = 7.0619(13)$ |
| | $b = 8.7085(16)$ |

|  | $c = 6.5562(11)$ |  |  |
|---|---|---|---|
|  | $\beta = 113.850(11)°$ |  |  |
| Volume, Å$^3$ | $V = 368.77(11)$ |  |  |
| Z | 4 |  |  |
| Density (calculated), g/cm$^3$ | 3.531 |  |  |
| Absorption coefficient, cm$^{-1}$ | $0.328 + 0.040·\lambda$ | / | None |
| F(000) | 137.7 | / | 384 |
| Crystal size, mm$^3$ | 6 × 6 × 17 | / | 1 × 1 × 1 |
| Number of reflections* | 17447 | / | 14819 |
| Data / restraints / parameters | 32267 / 0 /85 |  |  |
| Goodness-of-fit on $F^2$ | 7.78 | / | 3.23 |
| Final $R_1/wR_2$ indices | 0.0828 / 0.2353 | / | 0.0754/0.1248 |

*Outliers having |F$^2$ (obs) -F$^2$ (calc) |> 40.0 * σ(F$^2$ (obs)) were omitted from the refinement due to contribution from a second smaller crystallite.

Table 2: Structural parameters

| Atom | x | y | z |
|---|---|---|---|
|  | $U_{11}$ | $U_{22}$ | $U_{33}$ |
|  | $U_{12}$ | $U_{13}$ | $U_{23}$ |
| Ca | 0.24148(3) | 0.91837(2) | 0.75127(3) |
|  | 0.01937(7) | 0.00515(5) | 0.00712(5) |
|  | 0.00052(4) | 0.00034(4) | -0.00006(3) |
| Ti | 0.51469(2) | 0.75438(1) | 0.24959(2) |
|  | 0.00506(3) | 0.00490(4) | 0.00539(3) |
|  | -0.00004(2) | 0.00221(2) | 0.00004(2) |

| Si | 0.74843(3) | 0.93254(3) | 0.74902(3) |
| --- | --- | --- | --- |
| | 0.00507(4) | 0.00368(5) | 0.00423(4) |
| | -0.00005(3) | 0.00202(3) | -0.00009(4) |
| O(1) | 0.74957(3) | 0.82114(3) | 0.24948(4) |
| | 0.00536(4) | 0.00580(6) | 0.01109(6) |
| | -0.00008(4) | 0.00457(4) | 0.00006(6) |
| O(2) | 0.90973(3) | 0.81655(3) | 0.93419(3) |
| | 0.00892(6) | 0.00721(7) | 0.00552(6) |
| | 0.00188(4) | 0.00164(4) | 0.00118(5) |
| O(3) | 0.38313(3) | 0.96132(3) | 0.14756(3) |
| | 0.00812(5) | 0.00523(6) | 0.00721(6) |
| | 0.00177(4) | 0.00409(5) | 0.00084(5) |
| O(4) | 0.91186(3) | 0.31620(3) | 0.43642(3) |
| | 0.00873(6) | 0.00732(7) | 0.00543(6) |
| | 0.00160(4) | 0.00184(4) | 0.00116(5) |
| O(5) | 0.38073(3) | 0.46026(3) | 0.64715(3) |
| | 0.00814(5) | 0.00495(6) | 0.00719(6) |
| | 0.00154(4) | 0.00417(4) | 0.00059(5) |

To model the diffuse scattering, a Monte-Carlo simulation incorporating a simple ball-and-springs model connecting nearest and second-nearest neighbours was used with the atomic positions constrained to the values from the average structure. This corresponds to simulating diffuse scattering arising from thermal motion of the atoms. Whilst a detailed phonon dynamics cannot be obtained from this, it serves as a starting point to classify the origin of the diffuse scattering, as the mechanically simple model offers an intuitive physical interpretation. Classical harmonic springs were connected between nearest-neighbour cation and oxygen ions, such as Ti-O, Si-O and Ca-O. Additional springs were used to connect O-O neighbours within coordination polyhedra to prevent low-energy shearing modes from occurring that might distort the simulated crystal away from the average structure. The following energy expression was used to accept or reject a Monte-Carlo move corresponding to a random displacement of a randomly chosen atom [13]:

$$E = \sum_{i=1}^{n_{contacts}} k_i \cdot (d_i - d_0)^2$$

where $k_i$ is the force constant connecting the chosen atom with its neighbour $i$, $d_i$ and $d_0$ the current and equilibrium distance between this atom pair and the summation extends over all the contacts for the chosen atom. The usual Metropolis algorithm was used to accept/reject moves using the Boltzmann criterion with respect to $k_B T$ which was set to 1.0 [14]. The model crystal comprised 64 × 64 × 64 unit cells and computations were carried out using a sequential code on the CPU as well as a parallel code on an NVIDIA GTX285 graphics card. A method similar to [15] was implemented leading to an order of magnitude speed increase over the serial code. Diffraction patterns were computed using the program *diffusegpu* [16]. In each case, 200 Monte-Carlo cycles were run after which the model crystal did not show any deviation from the target $U_{iso}$ of Ti as this appears nearly isotropic. The force-constant ratios were kept constant during a simulation. Good qualitative agreement was achieved for the following ratios: Ca-O:Ti-O:Si-O:O-O:M-M = 10:50:100:5:5. M-M here means metal-metal interpolyhedral contact vectors. The *effective* force constants reflect the bonding strength and rigidity of the corresponding contacts. Hence, the $TiO_6$ and $SiO_4$ units are rather rigid whilst Ca which occupies large cavities is more loosely bound.

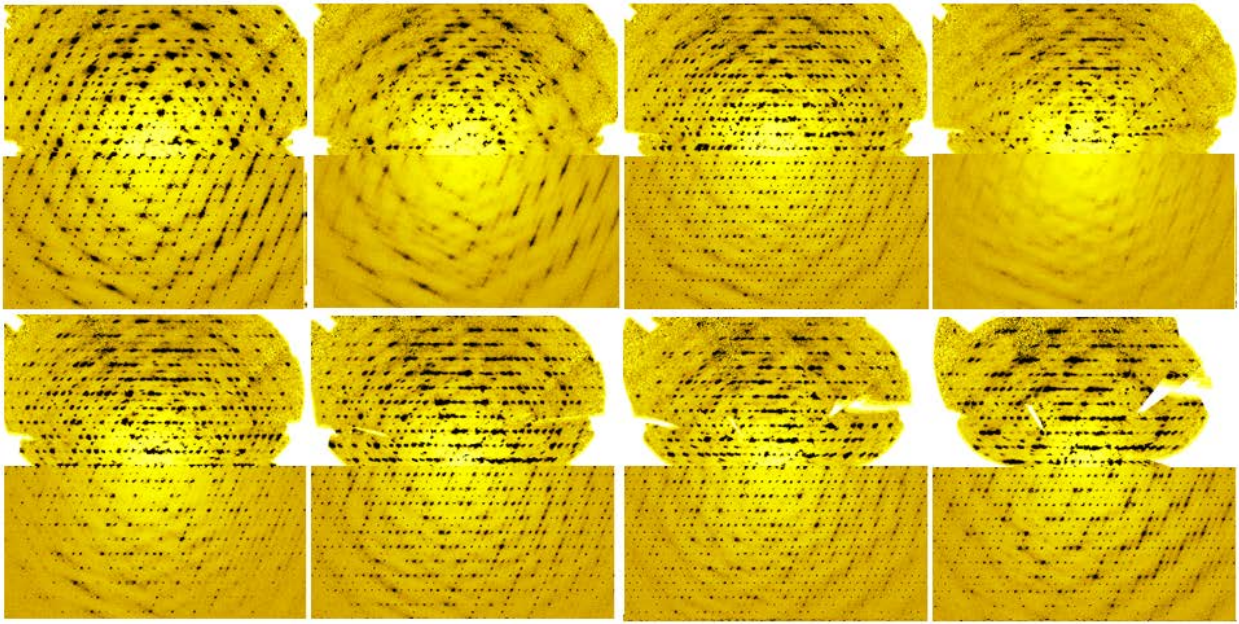

Figure 2 (Color online): Comparison of observed and calculated (upper and lower half in each image, respectively) neutron diffuse scattering, including Bragg peaks for the reciprocal space sections corresponding to, from left to right, upper row: ($h0l$), ($h,0.52,l$), ($h1l$), ($h,1.57,l$), lower row: ($h,2.09,l$), ($h,2.96,l$), ($h4l$), ($h,5.05,l$).

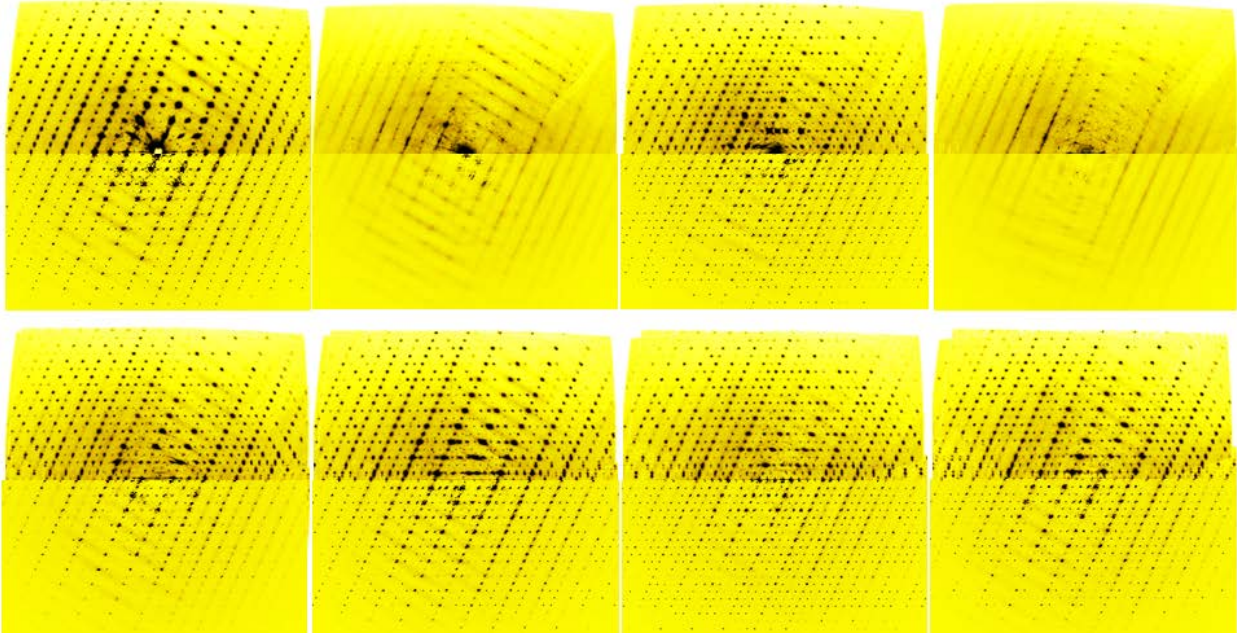

Figure 3 (Color online): Comparison of observed and calculated (upper and lower half in each image, respectively) X-ray diffuse scattering, including Bragg peaks for the reciprocal space sections corresponding to, from left to right, upper row: ($h0l$), ($h,0.52,l$), ($h1l$), ($h,1.57,l$), lower row: ($h,2.09,l$), ($h,2.96,l$), ($h4l$), ($h,5.05,l$).

## IV. Ab initio lattice dynamics calculations and comparison with diffuse scattering

DFPT calculations were performed using the plane-wave pseudopotential method as implemented in CASTEP [17]. Exchange and correlation were modelled in the Perdew-Burke-Enzerhof (PBE) approximation [18]. Pseudopotentials were of the norm-conserving variety generated using the Rappe-Rabe-Kaxiras-Joannopoulos (RRKJ) method [19], used with a plane-wave cutoff of 900 eV. A 4 × 3 × 4 Monkhorst-Pack grid was used to integrate electronic states across the Brillouin zone. These settings achieved convergence of forces to 0.015eV/Å and stress to 0.01 GPa. The crystal structure, including the cell parameters, was relaxed to equilibrium in response to an applied external pressure of 0 GPa. The lattice parameters converged to $a$ = 7.2028 Å, $b$ = 8.81017 Å, $c$ = 6.6778 Å, $\beta$ = 114.197°. Corresponding fractional coordinates of the atoms are given in Table 3. The deviations from the experimental unit cell are of the order of 2-3% which is typical of the behaviour of the PBE exchange correlation functional. Ab-initio lattice dynamics calculations were performed using the DFPT method [20] for a 4 × 3 × 4 grid of phonon wavevectors and the dynamical matrices were Fourier-interpolated to sample the Brillouin zone. This method includes all long-ranged electrostatic contributions and a correct description of the longitudinal optic–transverse optic (LO-TO) phonon splitting, which plays a strong role in the lattice dynamics of this system. Fourier-interpolation was used to compute phonons both on a fine grid for the diffuse scattering calculations and along high symmetry lines to generate phonon dispersion.

The resulting phonon dispersion curves are shown in Figure 4 (see also Supplemental Material at [*URL will be inserted by publisher*] for a compressed file containing details of the phonon dispersions). Given the 32 atoms per unit cell, a total of 3 × 32 phonon modes are expected. The phonon dispersions appear rather complex and for any given phonon mode, generally all the atoms participate to a varying degree in the corresponding real-space motions. Notably, a gap appears between about 600-700 cm$^{-1}$ with only two modes located inside the gap going from the Γ to the B point. Here, we follow the notation by Bradley and Cracknell to label the high-symmetry points [21]. These two modes are associated with the antiferroelectric Ti-O motion. The other modes can be broadly classified into complex tilting and bond-bending motions that have frequencies below the gap and high-frequency bond-stretching motions above the gap, e.g. breathing-type distortion of the octahedra and tetrahedra, though this classification is not strict. The lowest optic branches are dominated by the motion of Ca. One of the in-gap modes displays a discontinuity corresponding to a LO-TO splitting in agreement with infrared spectroscopy [2]. This particular mode involves a bond-stretching motion of the Ti along the *a* direction and will be discussed in more detail below.

Table 3: Structural parameters of CaTiSiO$_5$ after relaxation using CASTEP.

| Atom | x | y | z |
|------|---|---|---|
|      | $U_{11}$ | $U_{22}$ | $U_{33}$ |
|      | $U_{12}$ | $U_{13}$ | $U_{23}$ |
| Ca   | 0.240663 | 0.919353 | 0.750957 |
|      | 0.01948  | 0.00525  | 0.01286  |
|      | 0.00232  | -0.00806 | -0.00122 |
| Ti   | 0.513521 | 0.754565 | 0.249369 |
|      | 0.00375  | 0.00498  | 0.00426  |
|      | -0.00051 | 0.00022  | -0.00004 |
| Si | 0.748492 | 0.931417 | 0.748970 |
| | 0.00456 | 0.00368 | 0.00394 |
| | 0.00012 | -0.00040 | -0.00011 |
| O(1) | 0.749649 | 0.821802 | 0.249936 |
| | 0.00443 | 0.00588 | 0.00897 |
| | -0.00004 | 0.00191 | 0.00006 |
| O(2) | 0.906256 | 0.814541 | 0.933479 |
| | 0.00911 | 0.00748 | 0.00696 |
| | 0.00251 | -0.00258 | 0.00078 |
| O(3) | 0.382189 | 0.963517 | 0.149226 |
| | 0.00873 | 0.00582 | 0.00617 |
| | 0.00251 | 0.00088 | -0.00038 |
| O(4) | 0.908819 | 0.314581 | 0.435902 |
| | 0.00883 | 0.00742 | 0.00623 |
| | 0.00235 | -0.00212 | 0.00079 |
| O(5) | 0.380807 | 0.462910 | 0.649261 |
| | 0.00796 | 0.00536 | 0.00638 |
| | 0.00175 | 0.00096 | -0.00012 |

Xu and Chiang developed a formalism that allows to compute diffraction patterns corresponding to first-order thermal diffuse scattering given phonon polarisation vectors and eigenfrequencies [22]. For completeness, the key formulas are reproduced here. For one phonon mode, the structure factor, $F_j$, can be written as:

$$F_j(\vec{q}) = \sum_{n=1}^{n_{atoms}} \frac{b_n}{\sqrt{m_n}} \cdot e^{-\frac{1}{2}\sum_{k=1}^{3}\sum_{l=1}^{3} u_{kl,n} q_k q_l} \cdot \left(\vec{q} \cdot \vec{e}_{\vec{q},n,j}\right) \cdot e^{-i\vec{q}\cdot\vec{x}_n} \qquad (1)$$

These one-phonon structure factors are summed to yield the first order thermal diffuse intensity

$$I_{TDS}(\vec{q}) = C \cdot \sum_{j=1}^{n_{branches}} \frac{1}{\omega_{\vec{q}j}} \cdot coth\left(\frac{\hbar\omega_{\vec{q}j}}{2k_BT}\right) \cdot |F_j(\vec{q})|^2 \qquad (2)$$

where, $n_{atoms}$ is the number of atoms in the unit cell, $m_n$ the mass of the $n$-th atom, $b_n$ its neutron scattering length, $u_{kl,n}$ its thermal displacement parameter (see below), $\vec{x}_n$ its coordinates in an orthonormal crystal basis. $\vec{q}$ is the reciprocal lattice point and $\vec{e}_{\vec{q},n,j}$ is the phonon eigenvector at this point for atom $n$ and mode $j$. Eqs. (1) and (2) correspond to eqs. (34) and (35) in [22]. Note, that a belated corrigendum was added to equation (35) in the online version of the original paper. The occupancy factor for each phonon mode at a given temperature is obtained through the $coth\left(\frac{\hbar\omega_{\vec{q}j}}{2k_BT}\right)$ term in eq. (2), since DFPT calculations are performed at $T = 0$ K. The theoretically obtained anisotropic thermal displacement parameter was used for $u_{kl,n}$ as given in Table 3. These were calculated using the expression [23]:

$$u_{kl,n} = \frac{1}{2m_n} \int \frac{\hbar}{\omega_{\vec{q}}} \cdot \vec{e}_{\vec{q}k,n} \cdot \vec{e}^{\,*}_{\vec{q}l,n} \cdot coth\left(\frac{\hbar\omega_{\vec{q}}}{2k_BT}\right) \qquad (3)$$

where the integral is taken over the phonon frequencies and eigenvectors in the entire first Brillouin zone and '*' means complex conjugation. The indices $k$ and $l$ refer to the components $x, y, z$ and are the matrix elements of the thermal displacement tensor. The first Brillouin zone was divided into eight octants and expression (3) evaluated in each octant using a Gauss-Legendre integration and the contributions summed [24]. This scheme avoids potential problems due to the non-analytic behaviour of the LO modes at the Γ point. Rapid convergence was checked by employing a series of integrations with 4, 8, 16, 32, and 64 Gauss-Legendre support points in each dimension per octant. The numerical uncertainty is estimated to $2 \cdot 10^{-5}$ Å$^2$ which is less than the experimental standard deviation. Whilst the computed thermal parameters compare reasonably with the experiment, perfect agreement is not expected, since they are extrapolated from the 0 K DFT calculations neglecting anharmonicity and the phonon frequencies are likely offset by a few percent from the real ones.

The calculated patterns are compared with the neutron and X-ray data in Figures 5 and 6, respectively. As for the Monte-Carlo simulations corresponding to the simple Ball-and-Springs model, good visual agreement is obtained between the measured and calculated diffuse scattering patterns. Although not shown, it is noted that the simple Monte-Carlo simulation did not reproduce the thermal diffuse scattering (TDS) peaks underneath the Bragg peaks, as is expected and obtained from the acoustic phonons in the CASTEP calculations. However, the Monte-Carlo simulation reproduces the widely dispersed broad features and streaks.

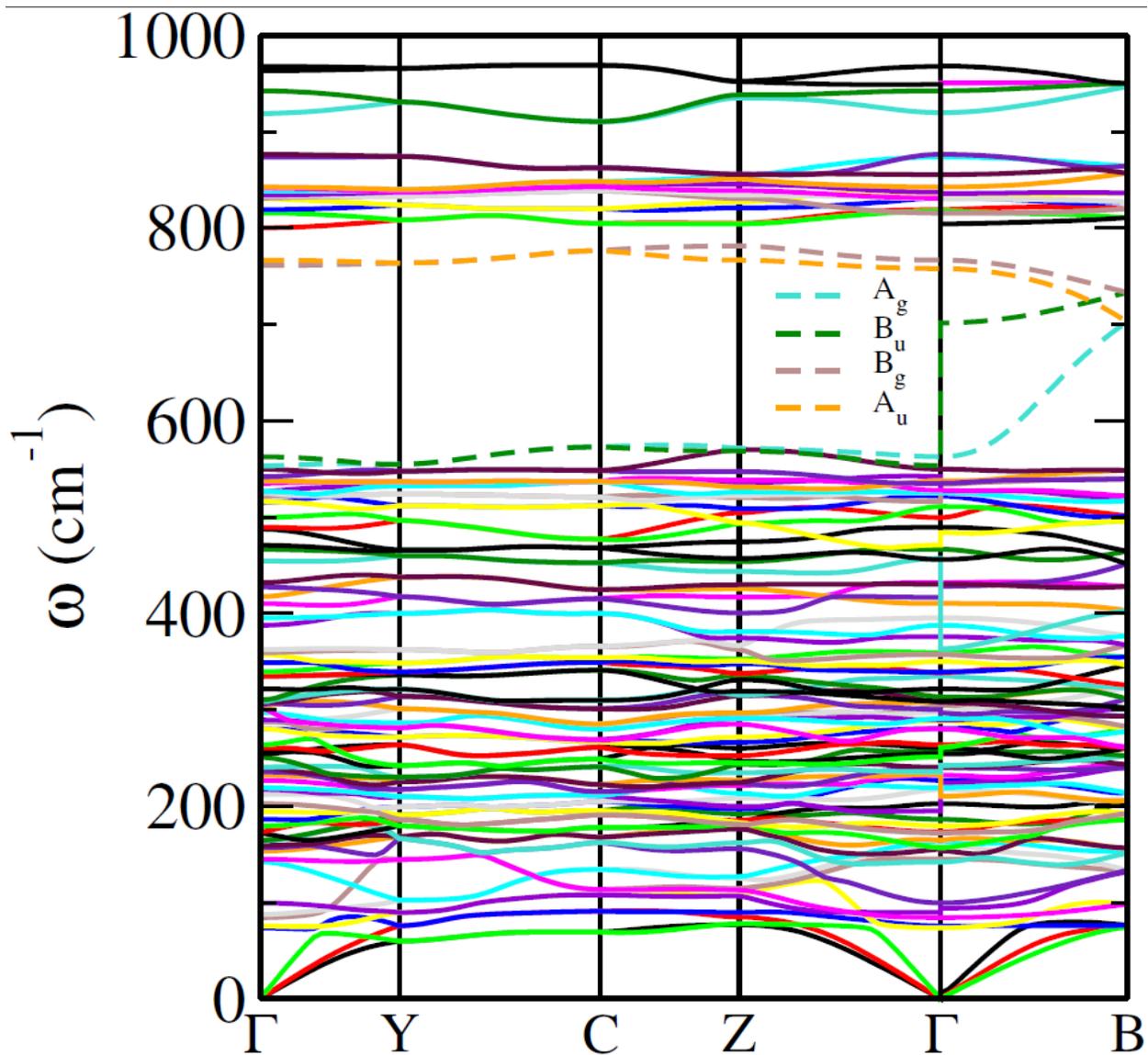

Figure 4 (color online): Calculated phonon dispersion curves for titanite along high-symmetry directions. The Brillouin zone labels are as follows from left to right along the horizontal axis: $\Gamma$ = (0, 0, 0), $Y$ = (0, 1/2, 0), $C$ = (0, 1/2, 1/2), $Z$ = (0, 0, 1/2), $\Gamma$ = (0, 0, 0), $B$ = (1/2, 0, 0) [21].

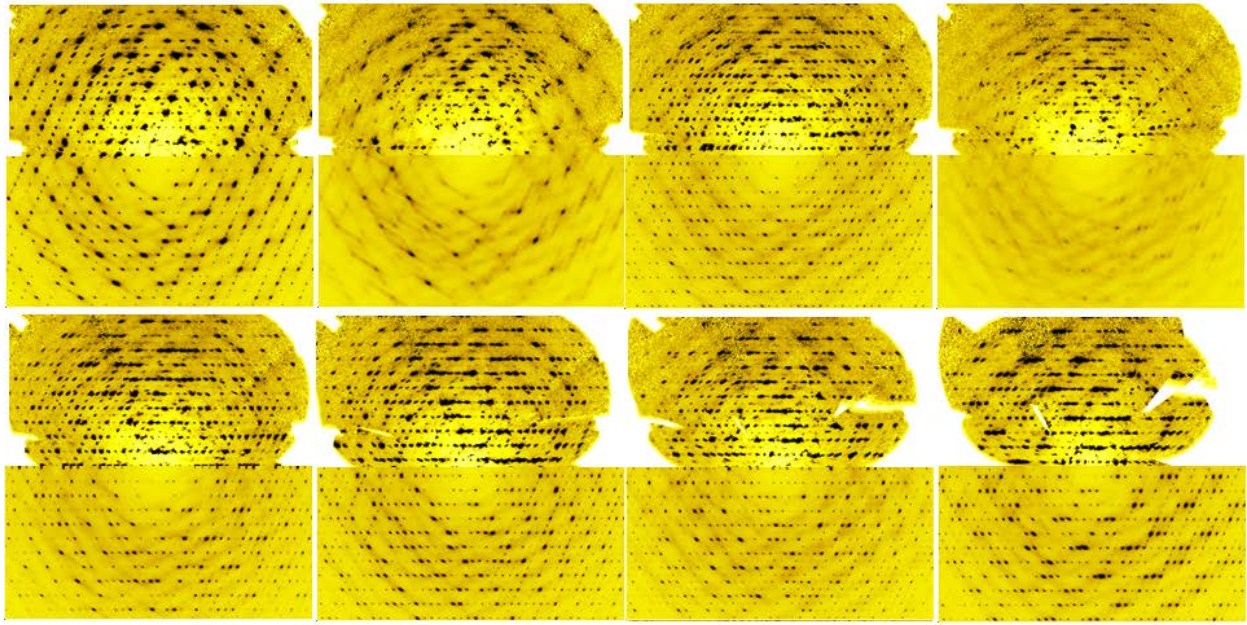

Figure 5 (Color online): Comparison of observed and calculated neutron diffuse scattering using DFPT and eqs. (1) and (2). The upper half in each plot are the data, lower half the pattern. Color scales of the computed images were adjusted to match the experimental data. Bragg peaks are not included in the theoretical patterns. The reciprocal space sections are the same as those in Figures 2 and 3.

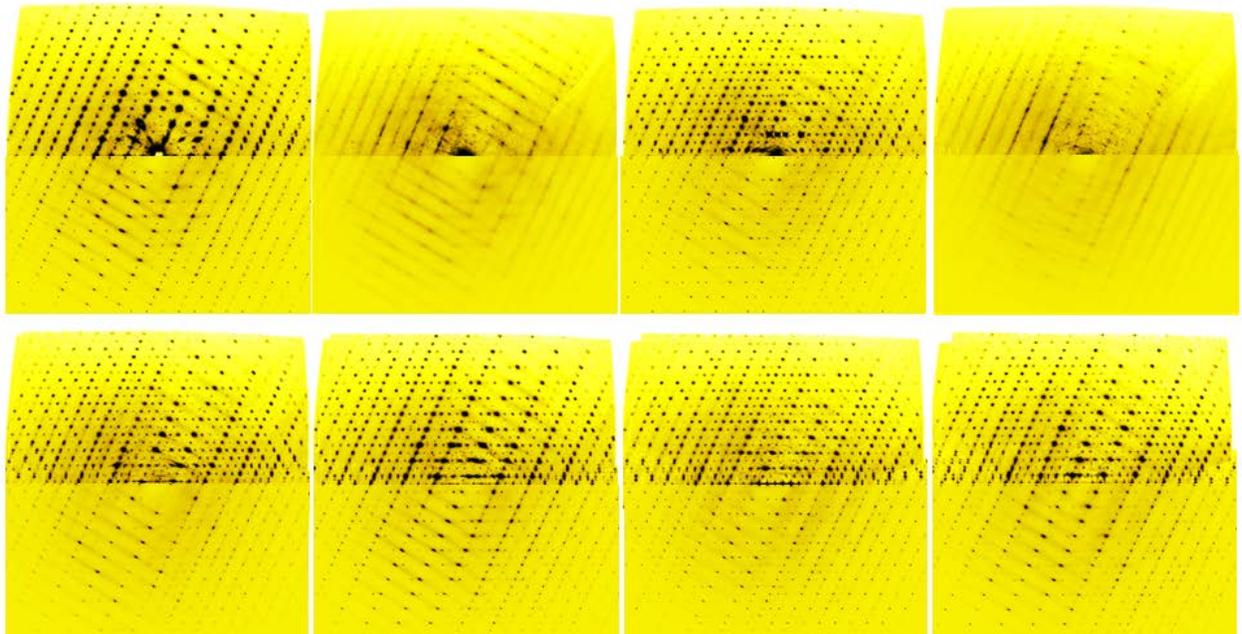

Figure 6 (Color online): Comparison of observed and calculated X-ray diffuse scattering using DFPT and eqs. (1) and (2). The upper half in each plot are the data, lower half the pattern. Bragg peaks are not included. Reciprocal space sections are the same as those in Figures 2 and 3.

## V. Discussion

The observed diffuse scattering is reproduced in great detail by the balls-and-springs model and the first order thermal diffuse scattering using ab-initio phonons. Examining the force-constant ratios from the former model allows us to conclude that the $TiO_6$ and $SiO_4$ polyhedra are rather rigid with Ca relatively loosely bound in its cage of surrounding oxygen. A comparison was made of two balls-and-springs model calculations that included and omitted second-nearest neighbours; this indicated that the inclusion of second-nearest neighbours is not essential to reproduce the diffuse scattering. More and complementary insight is gained from the ab-initio phonon calculations. The sharp peaks in the theoretical diffuse scattering maps originate essentially from the acoustic phonons. This is not surprising, since the weighting factor $1/\omega$ in Eq. (2) causes the TDS to peak for phonons having nearly zero frequency at the zone-center, coinciding with the location of Bragg peaks. The broad diffuse bands originate from the higher-frequency optic modes.

We would like to explore the sensitivity of the diffuse scattering with respect to the various phonon contributions. Figure 7 shows the reciprocal space maps ($h0l$) calculated using various frequency cutoffs. The corresponding X-ray data and the ($h$,0.52,$l$) layer are available as movies with each frame increasing the cutoff-frequency by 20 cm$^{-1}$ from the previous one. See Supplemental Material at [URL will be inserted by publisher]. This shows that the majority of features in the data are qualitatively explained using phonons up to about 200 cm$^{-1}$ with the phonons at higher energies merely adding intensity to existing diffuse features. Features appearing up to 80 cm$^{-1}$ are dominated by contributions from the acoustic branches and lowest-frequency optic modes with a dominant contribution from displacements of Ca, giving rise to the short horizontal streaks in the ($h$,0,$l$)-layer connecting two Bragg peaks (middle figure in top panel of Fig. 7). At higher frequencies, the assignment of particular phonon modes to diffuse features is no longer possible; these modes correspond to the polyhedral units moving approximately as rigid units in a concerted fashion involving bond-bending and tilt motions.

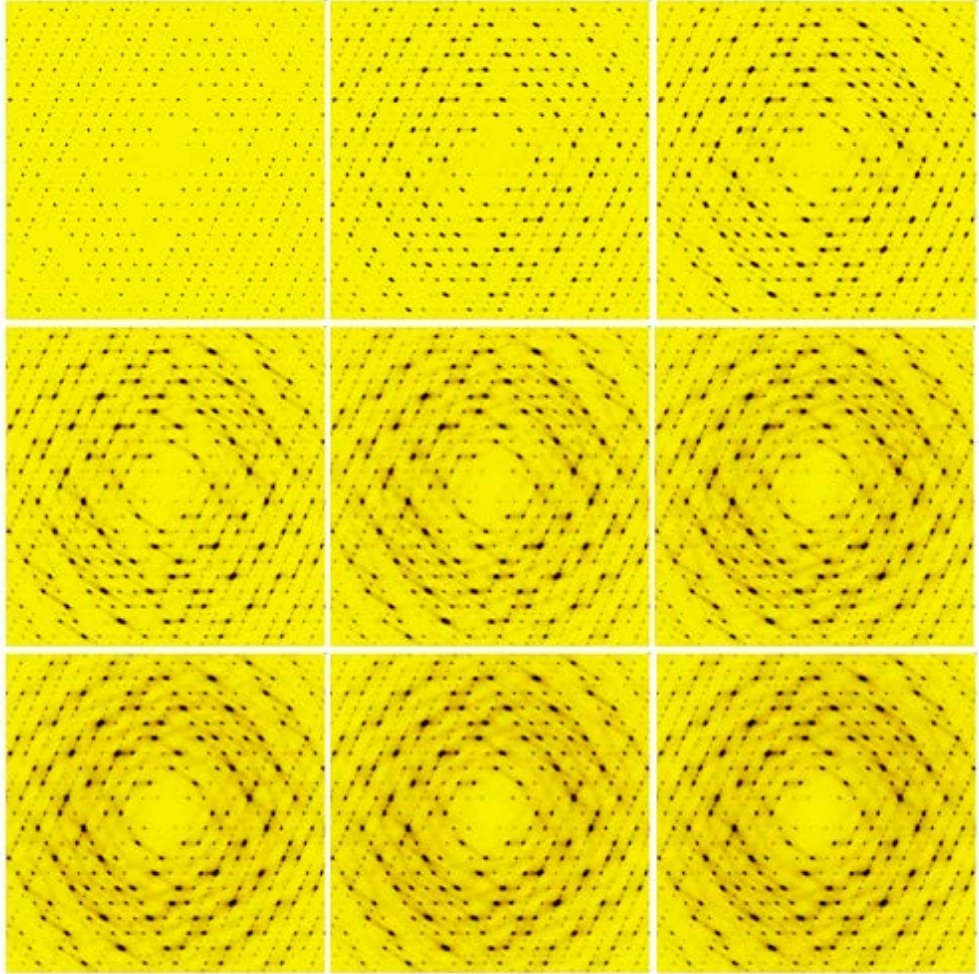

Figure 7 (Color online): ($h$,0,$l$) layer corresponding to the neutron data calculated for various frequency cutoffs of 40, 80, 120 cm$^{-1}$ (top row), 160, 200, 240 cm$^{-1}$ (middle row), and 280, 320, 360 cm$^{-1}$ (bottom row) going from left to right in each row.

Some of the high-frequency phonon modes show a peculiar behaviour. A large LO-TO mode splitting for the $B_u$ phonon in the gap between 600-800 cm$^{-1}$ is evident at the Γ point, where its frequency jumps from 562.5 cm$^{-1}$ to 701.4 cm$^{-1}$. This effect is illustrated in Figure 5 using dashed lines. This mode is one of a complex of four modes that involve a dominant Ti-O bond-stretching component along the $a$ direction, and which contribute strongly to the local microscopic polarizability. This complex arises from the symmetry of the displacements of those four oxygen atoms that form chains in the $a$ direction between the four Ti atoms in the unit cell. Of these, only the $B_u$ mode at 553 cm$^{-1}$ is fully symmetric with respect to both intra- and interchain oxygen motion. Consequently, only this $B_u$ mode can generate a macroscopic polarization for a Γ-point vibration, and it is the only one which displays a large LO-TO splitting. The $A_g$ mode at 563 cm$^{-1}$ is symmetric within each chain but out of phase with respect to inter-chain oxygen displacement. These chains are separated by 0.5 in the $c$ direction, and consequently the branch of symmetry $A_g$ at $\Gamma$ terminates at the *B* point (1/2,0,0) with in-phase O displacement. This motion also generates large internal electric fields and is responsible for the large dispersion of this mode along the $\Gamma$-*B* line. The $A_u$ mode at 758 cm$^{-1}$ corresponds to displacements that are also antisymmetric between chains, and the branch that extends from this mode also displays strong dispersion along $\Gamma$-*B*; this branch becomes degenerate at *B* with the branch that extends from the $A_g$ mode. The $B_g$ mode at 767 cm$^{-1}$ becomes degenerate at *B* with the termination of the branch that extends from the $B_u$ mode at $\Gamma$. The LO-TO splitting of the in-phase $B_u$ mode would be associated with any ferroelectic behaviour (as has been observed in piezoelectric crystals [25] and ferroelectrics [26] and references therein) but any of the other three modes could generate antiferroelectric local polarization. Three other $B_u$ modes with large LO-TO splittings of over 100 cm$^{-1}$ are present but are characterised by oxygen displacements predominantly perpendicular to the Ti-O chains and are therefore unlikely to contribute to macroscopic ferroelectricity.

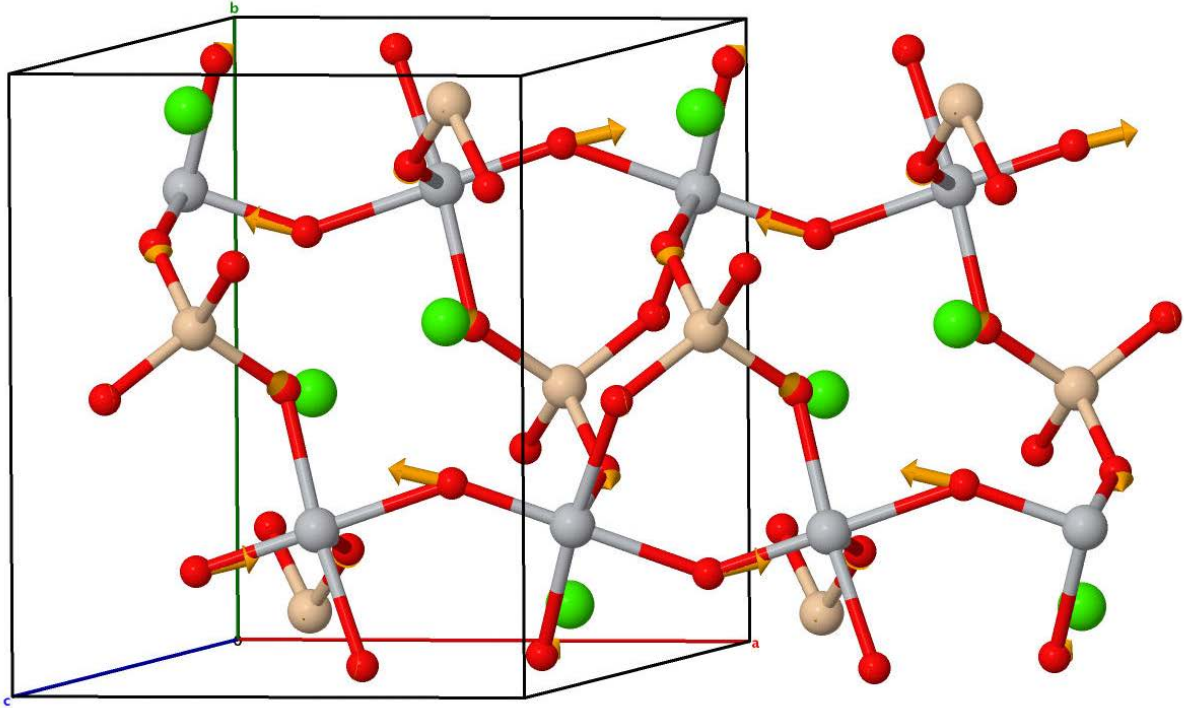

Figure 8 (Color online): Depiction of the real-space motion of the $B_u$ mode at 562.5 cm$^{-1}$ viewed roughly perpendicular to the (*a,b*)-plane. The coloring scheme is as follows: Ca (green), Ti (light grey), Si (beige), O (red). Arrows indicate the direction of motion.

## VI. Summary and conclusions

In this study, results from a joint neutron and high-energy X-ray single-crystal diffuse scattering experiment on $CaTiSiO_5$ at room-temperature have been presented. Two different approaches are shown to reproduce the observed diffuse scattering: One is a simple balls-and-springs model and the other uses ab initio phonon calculations. Whilst rather different in their scope and atomic-level detail, both models confirm the diffuse scattering originating from thermal motion of the atoms. The ball-and-springs model allows for a simple mechanical view of the system. The first-order TDS maps using the ab initio calculations reproduce many details and allow for a first insight into the lattice dynamics, given the lack of detailed experimental data on the phonon-dispersion relations away from the zone-center. As a word of caution though, the TDS features are qualitatively reproduced using the modes with a frequency cutoff of about 200 cm$^{-1}$ with the remainder of the phonons contributing additional intensity but no new diffuse features. A large LO-TO mode splitting is observed resulting in a large frequency shift for a $B_u$ mode at 562 cm$^{-1}$ to 701.4 cm$^{-1}$, respectively. Whilst generally involving all the atoms, this mode has a dominant component coming from Ti-O bond-stretching and, thus, the mode-splitting is related to the polarisability of the Ti-O bonds along the chain direction. Similar mode-splitting is observed in piezo- and ferroelectric materials. The combination of ab initio lattice dynamics and large scale reciprocal-space survey of the diffuse scattering sets a baseline from which to explore changes in the high-temperature phase and separate local structural changes from thermal effects.

## Acknowledgements

We would like to thank the late Prof. G. Marsaglia for advice on random number generators used in the Monte-Carlo simulations, in particular the parallel part, A. Skalski for assistance in crystal growth, and Dr. C. K. D. Stock for stimulating discussions. Experiments at the ISIS Neutron and Muon Source were supported by a beamtime allocation from the Science and Technology Facilities Council. Computing resources for DFPT calculations were provided by STFC's e-science facility, SCARF and by EPSRC grant EP/F037481/1 on the HECTOR supercomputer.